\documentclass[useAMS,usenatbib]{mn2e}
\usepackage{graphicx}
\usepackage{caption}
\usepackage{subcaption}
\usepackage{amssymb}
\usepackage{hyperref}

\newcommand{\ion}[2]{#1$\,${\small \MakeUppercase{\romannumeral #2}}}
\newcommand{\OI}{[\ion{O}{1}]\,$\lambda\lambda 6300$, 6364}
\newcommand{\Ca}{[\ion{Ca}{2}]\,$\lambda\lambda 7291$, 7324}
\newcommand{\Mg}{\ion{Mg}{1}]\,$\lambda 4571$}
\newcommand{\kms}{km\,s$^{-1}$}

\title[Nebular Spectroscopy of SN~2011dh]
 {Nebular Spectroscopy of the Nearby Type IIb Supernova~2011dh}

\author[Shivvers et al.]
 {Isaac Shivvers,$^{1\dagger}$ Paolo Mazzali,$^{2,3}$
 Jeffrey M. Silverman,$^{4,5}$ J\'anos Boty\'anszki,$^6$
 \newauthor
S. Bradley Cenko,$^{1,7}$ Alexei V. Filippenko,$^1$
Daniel Kasen,$^{1,6,8}$ Schuyler D. Van Dyk,$^9$
\newauthor
Kelsey I. Clubb$^1$ \\
$^1$Department of Astronomy, University of California, Berkeley, CA 94720-3411, USA \\
$^2$Astrophysics Research Institute, Liverpool John Moores University, Liverpool, UK \\
$^3$Max-Planck-Institut f\"ur Astrophysik, Karl-Schwarzschildstr.~1, D-85748 Garching, Germany \\
$^4$Department of Astronomy, University of Texas, Austin, TX 78712, USA \\
$^5$NSF Astronomy and Astrophysics Postdoctoral Fellow \\
$^6$Department of Physics, University of California, Berkeley, CA 94720, USA \\
$^7$Astrophysics Science Division, NASA/Goddard Space Flight Center, Mail Code 661, Greenbelt, MD, 20771, USA \\
$^8$Nuclear Science Division, Lawrence Berkeley National Laboratory, Berkeley, CA 94720, USA \\
$^9$Spitzer Science Center/Caltech, Mailcode 220-6, Pasadena, CA 91125}

\date{Accepted to MNRAS; 2013 September 25.  \\
$^\dagger$Email: ishivvers@astro.berkeley.edu}

\begin{document}
\maketitle

\begin{abstract}

We present nebular spectra of the nearby Type IIb supernova (SN) 2011dh taken between 201 and 
678\,days after core collapse.  At these late times, SN~2011dh exhibits strong emission lines including a broad and persistent H$\alpha$
feature.  New models of the nebular spectra confirm that the progenitor of SN~2011dh was a low-mass giant ($M \approx 13$--15\,M$_{\sun}$)
that ejected $\sim$\,0.07\,M$_{\sun}$ of $^{56}$Ni and $\sim$\,0.27\,M$_{\sun}$ of oxygen at the time of explosion, consistent with the recent disappearance of 
a candidate yellow supergiant progenitor. We show that light from the SN location is dominated by the fading SN at very late times ($\sim$\,2\,yr)
and not, for example, by a binary companion or a background source.  We present evidence for interaction between the expanding SN blastwave 
and a circumstellar medium at late times and show that the SN is likely powered by positron deposition $\gtrsim$\,1\,yr after explosion.
We also examine the geometry of the ejecta and show that the nebular line 
profiles of SN~2011dh indicate a roughly spherical explosion with aspherical components or clumps. 

\end{abstract}


\section{Introduction}

Type IIb supernovae \citep[SNe;][]{Woosley:1987,Filippenko:1988} are a relatively rare class of core-collapse 
supernova (SN), constituting only $\sim$\,7\% of all SNe \citep{Li:2011cl}. Like
other SNe~II, they show strong hydrogen features in their early-time spectra, yet within only
a few weeks after core collapse the H fades and the spectra of SNe~IIb most closely resemble those
of stripped-envelope SNe~Ib \citep[for a review of the spectral classification of SNe, see][]{Filippenko:1997dz}.
SNe~IIb therefore represent a transitional class of core-collapse SNe with only partially stripped envelopes. Exactly what process 
removes most (but not all) of their hydrogen envelope is still an open question, though interaction with a binary companion increasingly appears
to be the most likely answer.  

Thus far, there have been only a handful of
nearby and intensely studied SNe~IIb, including SN~2008ax \citep[$\sim$\,9.6\,Mpc; e.g.,][]{Chornock:2011dj}, 
SN~2001ig \citep[$\sim$\,11.5\,Mpc; e.g.,][]{Silverman:2009fo},
SN~2003bg \citep[$\sim$\,21.7\,Mpc; e.g.,][]{Hamuy:2009jf,Mazzali:2009dd},
and SN~1993J \citep[$\sim$\,3.69\,Mpc; e.g.,][]{Filippenko:1993jt,Matheson:2000}. 
SN~2011dh in M51 ($\sim$\,8.05\,Mpc; see \S\ref{sec:dist} below) has become another nearby and very well-observed
example of this unusual class of SN.

In early June 2011, SN~2011dh (also known as PTF11eon) was independently discovered within $\sim$\,1\,day of core collapse by 
several amateur astronomers \mbox{\citep{Griga:cbet}} and the
Palomar Transient Factory collaboration \citep[PTF;][]{PTF:a,PTF:b,Arcavi:2011gb}.
The SN is apparent in an image taken by A.~Riou of France on May 31.893
(UT dates are used throughout), while a PTF image taken May 31.275 does not detect a source down
to a 3$\sigma$ limiting magnitude of $m_g = 21.44$ \citep{Arcavi:2011gb}. These observations most likely bracket
the time of explosion, and for this paper we assume an explosion date of May 31.5.  
After discovery, a spectrum was promptly obtained by \citet{Silverman:atel}, and a possible progenitor star was first identified in archival {\it Hubble
Space Telescope (HST)} images by \citet{Li:atel:a}.

\citet{Maund:2011ez} and \citet{VanDyk:2011jj} confirmed the identification of the likely progenitor
star in {\it HST} images through ground-based adaptive optics imaging of the SN, measuring a spatial coincidence of 
the {\it HST} source and the SN to within 23 and 7\,mas, respectively.  Both reported that the source in the {\it HST} images
has a spectral energy distribution consistent with a single star: a yellow (mid-F) supergiant with an extended
envelope ($R \approx 200\,{\rm R}_{\sun}$), a temperature of $\sim$\,6000\,K, and a mass of 13--18\,M$_{\sun}$.
However, \citet{VanDyk:2011jj} expressed doubt that the yellow supergiant (YSG)
is the true progenitor of SN~2011dh, instead preferring a scenario with a faint and compact progenitor as a binary
companion to the YSG. This was largely motivated by the results of \citet{Arcavi:2011gb}, who
favored a compact ($10^{11}$\,cm) binary companion based on the rapidity of 
the shock breakout and the relatively cool early photospheric temperature.
\citet{Soderberg:2012gt} supported this interpretation with radio and
X-ray observations, estimating the progenitor size to be $\sim$\,$10^{11}$\,cm through modeling of the cooling
envelope. In this compact-star scenario, the progenitor of SN~2011dh was theorised to be a faint Wolf-Rayet
star with a zero-age main sequence mass $\gtrsim$\,25\,M$_{\sun}$ and a history of mass loss 
through vigorous winds.

\citet{Bersten:2012ic} disagreed; their hydrodynamical models suggested that an extended progenitor
was required to produce the early-time light curve, at odds with the analytic relation used by \citet{Arcavi:2011gb},
originally from \citet{Rabinak:2011ku}.  \citet{Bersten:2012ic} found that a progenitor with a zero-age main sequence
mass of 12--15\,M$_{\sun}$ and a radius $\sim$\,200\,R$_{\sun}$ was consistent
with the early-time light curve and photospheric temperature, and showed that any model with a zero-age main sequence mass $\gtrsim$25\,M$_{\sun}$
(i.e., a Wolf-Rayet star) was strongly disfavoured.  \citet{Benvenuto:2012gj} presented a model of a possible
binary progenitor scenario with a $\sim$\,16\,M$_{\sun}$ YSG primary star losing material to a much fainter
$\sim$\,10\,M$_{\sun}$ companion (undetectable in the pre-explosion {\it HST} images).

In addition, \citet{Murphy:2011em} argued that the mass of the SN~2011dh progenitor must be either $13^{+2}_{-1}$\,M$_{\sun}$ 
or $>$\,29\,M$_{\sun}$, based upon an analysis of the star-formation history of the SN's environment.  Star formation in 
the vicinity of the SN overwhelmingly occurred in two discrete bursts at $<$\,6 and $17^{+3}_{-4}$\,Myr; the zero-age
main sequence mass of the SN is constrained by assuming the star is associated with one of those events, taking into
account errors due, for example, to uncertain late-stage stellar evolution and mass loss.  This result is consistent with
the YSG progenitor scenario. Throughout 2012 other authors presented further panchromatic observations, some of which favoured
a compact progenitor while others suggested an intermediate or extended progenitor, emphasising the
need for a definitive progenitor identification
\citep[e.g.,][]{Krauss:a,Bietenholz:a,Campana:a,Horesh:a,Sasaki:a}.

The desired identification was provided by \citet{VanDyk:2013tp}, who reported that the YSG progenitor candidate had disappeared from new {\it HST} images.
Specifically, at an age of $\sim$\,641\,days SN~2011dh had faded down to 1.30 and 1.39\,mag fainter than the YSG progenitor
in the {\it HST} Wide Field Camera 3 (WFC3) $F555W$ and $F814W$ passbands, respectively.  This result is corroborated by 
\citet{Ergon:2013td}, who report a significant decline in the $B$, $V$, and $r$-band fluxes between pre-explosion imaging
of the YSG progenitor and imaging of the SN at 600+\,days past explosion.
These results clearly point toward the extended YSG progenitor found in archival {\it HST} images as
the progenitor star of SN~2011dh.

In this paper, we present six new spectra of SN~2011dh taken between 201 and 678\,days after core collapse, in the nebular phase of its evolution.
During the nebular phase, the SN ejecta are optically thin and we can directly observe the products of explosive nucleosynthesis without reprocessing
through a photosphere.  Our very late-time spectra show that the flux observed by \citet{VanDyk:2013tp} and \citet{Ergon:2013td} was produced primarily
by the fading SN and not a stellar source.
We present models of the nebular emission spectra and detailed analyses of the line profiles and the late-time flux energetics, 
providing constraints on the progenitor's mass and composition and the geometry of the explosion.
We describe our observations and data-reduction procedure in \S\ref{sec:obs}, present our spectra and analysis in \S\ref{sec:ana}, 
discuss our model spectra in \S\ref{sec:models}, and conclude in \S\ref{sec:conc}.

\section{Observations and Data Reduction}
\label{sec:obs}

\subsection{Spectroscopy}

Following its discovery in early June 2011, we began an extensive spectroscopic monitoring campaign of SN~2011dh.
Some of our early-time spectra from the Lick and Keck Observatories (including a spectrum obtained only 2.4\,days after explosion)
have already been presented by \citet{Arcavi:2011gb}, and other groups have published their own spectra 
\citep{Marion:2013vr,Ergon:2013td,Sahu:2013vd}. This study focuses on the nebular phase of SN~2011dh.

We collected spectra using both the Lick and Keck Observatories, moving to a larger aperture as the SN faded away.
We used the Kast double spectrograph on the Shane 3\,m telescope at Lick Observatory \citep{kast}, the Low Resolution Imaging
Spectrometer (LRIS) mounted on the 10\,m Keck I telescope \citep{lris},
and the DEep Imaging Multi-Object Spectrograph (DEIMOS) on the 10\,m Keck II telescope \citep{deimos}
to collect 3, 1, and 2 nebular spectra of SN~2011dh, respectively. Table \ref{tab:spec} summarises observing details
for these 6 spectra.  

\begin{table*}
\begin{minipage}{126mm}
\caption{Journal of spectroscopic observations}
\label{tab:spec}
\begin{tabular}{ l *5c }
\hline
UT Date & Day\footnote{Days since explosion (2011 May 31.5).} &
Instrument &  Range (\AA) &
Resolution\footnote{Approximate resolution (averaged over spectrum).} (\AA) &
Exposure time (s) \\
\hline
2011 Dec.  18 &  201   & Kast             & 3480--10114         & 7       & 2400 \\
2011 Dec.  24 &  207   & Kast             & 3428--10100         & 6       & 2100 \\
2012 Feb.  23 &  268   & Kast             & 3444--10166         & 6       & 3000 \\
2012 Apr.   29 &  334   & LRIS            & 3312--7360           & 3       & 600  \\
2013 Feb.  17 &  628   & DEIMOS     & 4500--9640           & 3       & 2400 \\
2013 Apr. 8     &  678   & DEIMOS      & 4450--9603          & 3       & 3600 \\
\hline
\end{tabular}
\end{minipage}
\end{table*}

\subsection{Data Reduction}
All observations were collected and reduced following standard techniques as described by \citet{Silverman:2012be}.
All spectra were taken with the slit oriented at the parallactic angle to minimise
flux losses caused by atmospheric dispersion \citep{Filippenko:1982br}. 
We use a low-order polynomial fit to arc-lamp observations to calibrate the wavelength scale, and we
flux calibrate each spectrum with a spline fit to standard-star spectra observed the same night at
a similar airmass. In addition, we have removed telluric absorption lines from all spectra. 
Upon publication, all raw spectra presented in this paper will be made available in electronic format on WISeREP
\citep[the Weizmann Interactive Supernova data REPository;][]{wiserep}.\footnote{\url{http://www.weizmann.ac.il/astrophysics/wiserep}}

\subsection{Distance}
\label{sec:dist}

The distance to M51 has been measured through several independent methods with significant scatter among
their results. We follow \citet{Marion:2013vr} and adopt $D=8.05 \pm 0.35$\,Mpc, an average of four of these measures
\citep{Tonry:ab,Tully:ab,Vinko:2012cw,Feldmeier:ab}.
All spectra have been deredshifted by M51's recession velocity, 600\,\kms\ \citep[$z = 0.002$, NED;][]{Rush}.  M51
is at very low redshift and so we neglect time-dilation effects due to cosmological expansion.
Both \citet{Arcavi:2011gb} and \citet{Vinko:2012cw} use high-resolution spectra to measure the reddening toward M51
using \ion{Na}{1} D absorption-line widths.  Both find the host-galaxy extinction to be negligible and the Milky Way
extinction to be consistent with values measured by \citet{dustmaps}: $E(B-V) = 0.035$\,mag. We deredden
all spectra by this value prior to analysis, using the reddening law of \citet{redlaw}
and assuming $R_V=3.1$.  Note that \citet{Ergon:2013td} adopted a slightly higher value of
$E(B-V) = 0.07^{+0.07}_{-0.04}$\,mag, corresponding to a $\sim$\,5--10\% difference in absolute flux 
level across the optical spectrum, not enough to significantly affect the discussion below.

\subsection{Absolute Flux Calibration}
\label{sec:fluxcal}

\begin{figure}
\centering
\includegraphics[width=0.5\textwidth]{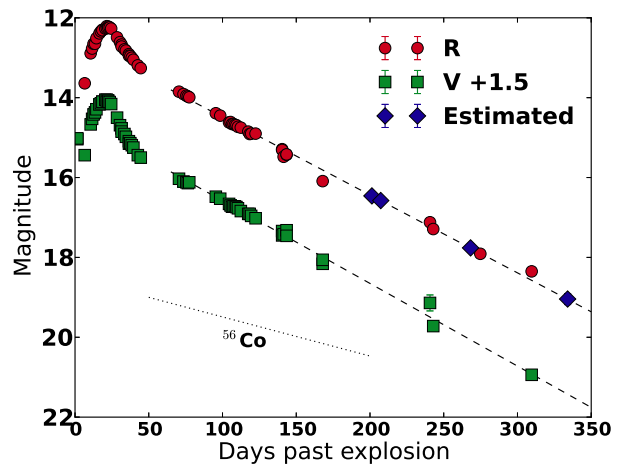}
\caption{Photometric decline of SN~2011dh from \citet{Tsvetkov:2012ty}, with our best-fit decline rates and the $^{56}$Co
               decay rate overplotted.  The blue (diamond) points are the estimated values used to flux calibrate our spectra between days 201 and 334.
               Note that error bars for most points are smaller than the plotted symbol. A colour version of this figure is available in the online journal.}
\label{fig:phot}
\end{figure}

Our observation techniques and data-reduction methods record the relative flux with high fidelity, but 
absolute flux calibrations are a persistent difficulty in long-slit spectroscopy.  Variations in atmospheric seeing
between flux-standard observations and science observations can result in varying amounts of flux falling out of the slit 
and spectral observations are often taken in less-than-photometric conditions with nonnegligible
(and possibly varying) levels of cloud cover.
Parts of our analysis (see \S\S\ref{sec:latespec}, \ref{sec:models}) require an absolute flux measure, however, so we
address this problem by flux calibrating our spectra to late-time photometry of SN~2011dh wherever possible.

\citet{Tsvetkov:2012ty} present $UBVRI$ light curves of SN~2011dh extending
to just over 300\,days; we assume a linear decay in $R$-band magnitudes beyond $\sim$\,70\,days and perform a
maximum-likelihood analysis to estimate the $R$ magnitude of SN~2011dh at the time each spectrum was taken.
We chose the $R$ band because of its relatively dense coverage and because several of the strongest nebular lines
([\ion{O}{1}], [\ion{Ca}{2}], \ion{Na}{1}, H$\alpha$) fall within the passband, making it a good tracer of the SN's decline. 
We match synthetic photometry of our 201--334\,day spectra to these values.
All synthetic photometry has been calculated with \verb|pysynphot| \citep{pysynphot}.
As shown in Figure~\ref{fig:phot}, we find an $R$-band decline rate of of $0.0195 \pm 0.0006$\,mag\,day$^{-1}$ and a $V$-band decline
rate of $0.0207 \pm 0.0009$\,mag\,day$^{-1}$ (reported errors are 68\% confidence levels; $\sim 1\sigma$).

A linear decay in magnitudes is a reasonable assumption so long as emission is primarily driven by
the radioactive decay of $^{56}$Co \citep[e.g.,][]{Colgate:1969jw,arnett:1996}.  It is common for SNe~Ib/IIb to display decline rates significantly faster than the 
$^{56}$Co $\rightarrow$ $^{56}$Fe rate (0.0098\,mag\,day$^{-1}$) -- a steep decline rate is reasonably interpreted as evidence for
a declining $\gamma$-ray trapping fraction in the ejecta (as the ejecta expand and the density drops, more of the $\gamma$-rays
produced by $^{56}$Co decay escape before depositing their energy).
The decline rate of SN~2011dh is slightly faster than those measured for both SN~1993J and SN~2008ax,
two well-understood SNe~IIb which had decline rates of 0.0157 and 0.0164\,mag\,day$^{-1}$, respectively \citep[fit to days $\sim$\,60--300;][]{Taubenberger:2011hx}.
See \S\ref{sec:latespec} for a comparison between these early-time nebular decline rates and the flux observed at very late
times ($>$\,600\,days).

We do not assume that the same decay law holds true out to our last two spectra, at 628 and 678\,days after core collapse.
Instead, we repeat the analysis described above using photometry from \citet{Ergon:2013td}, 
who report Nordic Optical Telescope (NOT) observations in $V$ at 601 and 685\,days.

\section{Analysis}
\label{sec:ana}

\begin{figure*}
\centering
\begin{subfigure}[b]{0.49\textwidth}
\centering
\includegraphics[width=\textwidth]{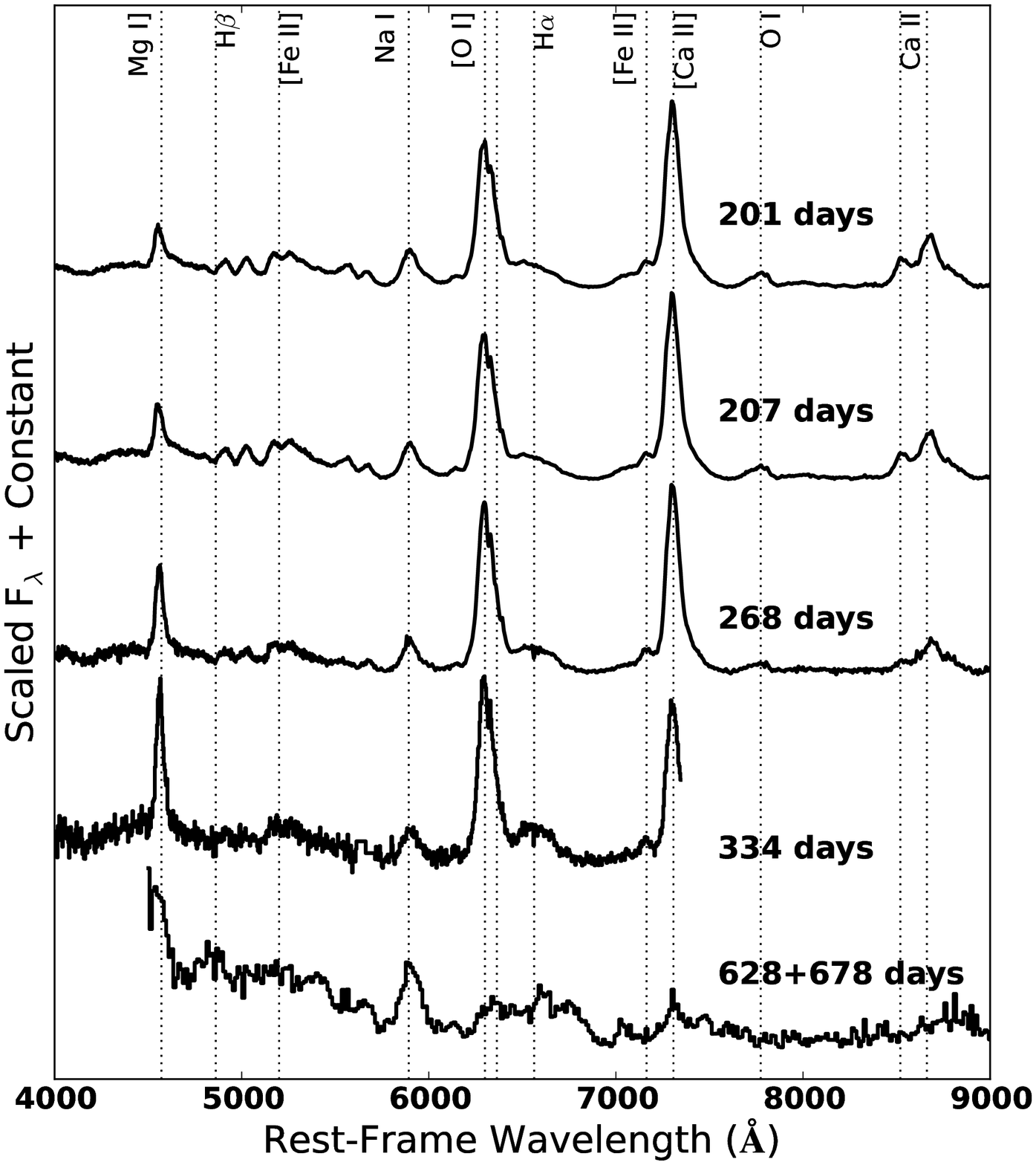}
\end{subfigure}
~
\begin{subfigure}[b]{0.49\textwidth}
\centering
\includegraphics[width=\textwidth]{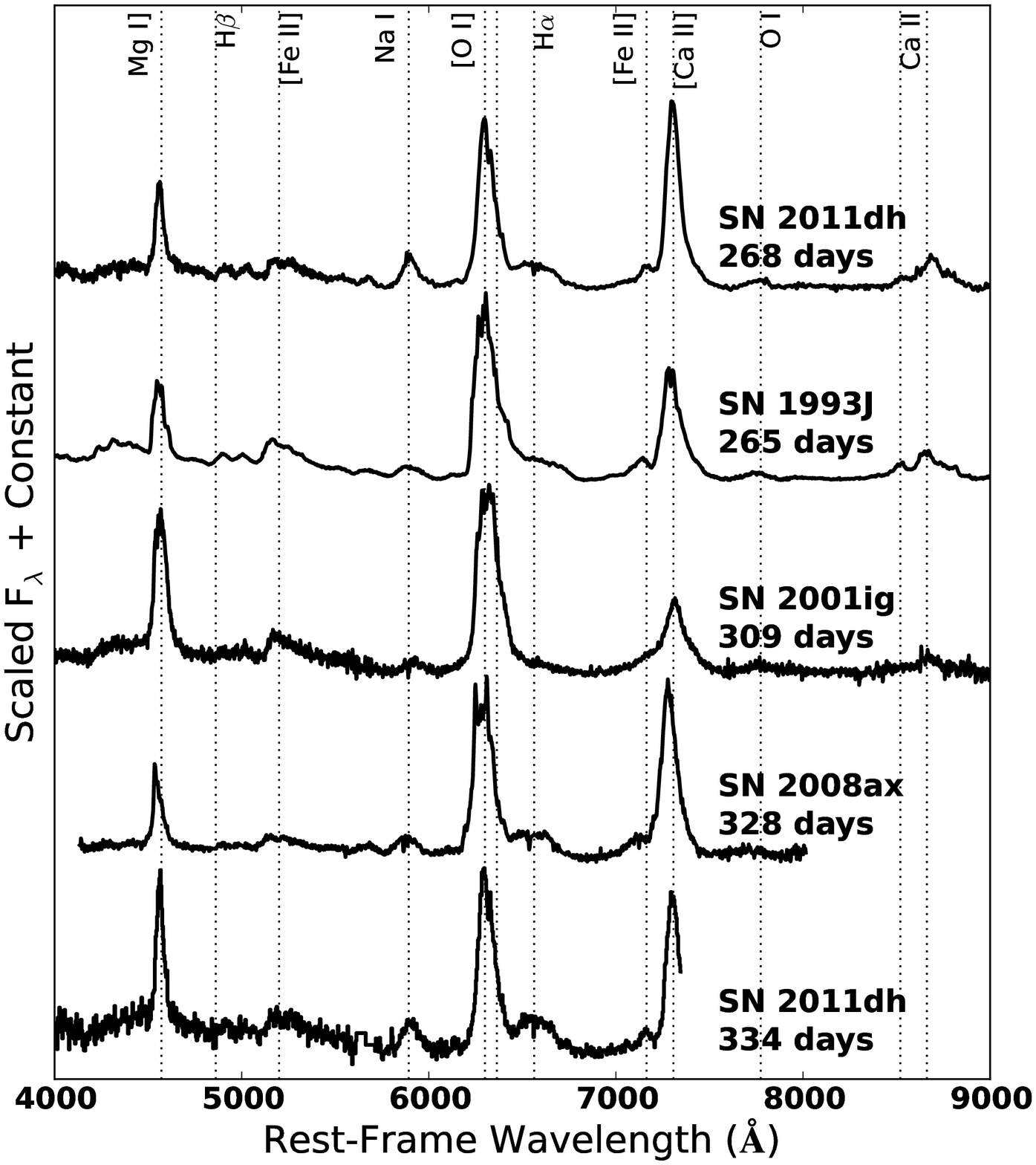}
\end{subfigure}
\caption{Nebular spectra of SN~2011dh (left) and comparison spectra of SN~1993J, SN~2001ig, and SN~2008ax (right).
               All spectra have been deredshifted.  All displayed SNe are at low redshift and time-dilation effects are negligible;
               listed phases are days since explosion in Earth's reference frame.
               The spectra of SN~2011dh from 628 and 678\,days have been coadded and rebinned to increase the signal-to-noise ratio (S/N). 
               The SN~1993J spectrum is from \citet{Filippenko:1994is} and \citet{Matheson:2000ff}, the SN~2001ig spectrum is from \citet{Silverman:2009fo},
               and the SN~2008ax spectrum is from \citet{Milisavljevic:2010cl}.}
\label{fig:compspec}
\end{figure*}

By 201\,days past explosion SN~2011dh was well into the nebular phase, with 
a spectrum dominated by strong emission lines and little or no continuum.  Figure \ref{fig:compspec} shows our complete spectral sequence
of SN~2011dh in the nebular phase with spectra from 201 to 678\,days after explosion and a few prominent lines identified, 
and compares the spectra of SN~2011dh to those of a few other prominent SNe~IIb at comparable epochs.

Throughout the first year after explosion the nebular spectra of SN~2011dh are dominated by strong \OI~and
\Ca~emission lines, alongside a strong \Mg~emission line and persistent \ion{Na}{1}~D and H$\alpha$ lines.
Table \ref{tab:flux} lists relative line strengths of several prominent lines in the early
nebular phase. We measured these fluxes by subtracting a local linear continuum and integrating
over each line. Note that the continuum here is not from the photosphere of the SN, but rather is 
likely a mixture of blended lines, producing a sort of pseudocontinuum.
Also, note that this type of integrated flux measurement is by no means exact due to line blending
and the approximated local continuum, but care was taken to treat each line similarly and
these measures should accurately portray the relative-flux trends.

The relative flux of [\ion{Ca}{2}] and [\ion{O}{1}] has been shown to be a useful indicator of progenitor core mass, with
smaller [\ion{O}{1}]/[\ion{Ca}{2}] ratios generally indicative of a less massive helium core at the time of explosion  \citep[e.g.,][]{Fransson:1989ie,Jerkstrand:2012gu}.
SN~2011dh displays an [\ion{O}{1}]/[\ion{Ca}{2}] ratio significantly smaller than that in both SN~1993J and SN~2001ig.
The ratio is similar to that in SN~2008ax, which also displayed a similar upward trend throughout the nebular phase
\citep{Silverman:2009fo,Filippenko:1994is,Chornock:2011dj}.
It therefore appears that SN~2011dh's progenitor He core mass was relatively close to that
of SN~2008ax and significantly less than that of both SN~2001ig and SN~1993J.  See \S\ref{sec:models} for a more
thorough analysis.

\begin{table*}
\begin{minipage}{130mm}
\caption{Integrated line fluxes relative to \Ca}
\label{tab:flux}
\begin{tabular}{ *8c }
\hline
Day\footnote{Days since explosion (2011 May 31.5).} & 
[\ion{Ca}{2}] $+$ \ion{Fe}{2}\footnote{\Ca~and \ion{Fe}{2}\,$\lambda 7155$ are difficult to deblend, so we present the integrated flux of both.  
                                                                  The contribution from \ion{Fe}{2} is, however, very much smaller than [\ion{Ca}{2}] (see Fig.~\ref{fig:compspec}).} &
\ion{Mg}{1}] & \ion{Na}{1}~D &
\ion{Ca}{2}\footnote{The \ion{Ca}{2} near-infrared triplet.} & \ion{O}{1}\,$\lambda 7774$ &
H$\alpha$\footnote{To measure the blended [\ion{O}{1}] and H$\alpha$ lines we assume H$\alpha$ is symmetric about the rest wavelength (6563\,\AA).
                                    We report the H$\alpha$ flux as twice the value obtained by integrating from the red continuum to 6563\,\AA.} &
\OI\footnote{The [\ion{O}{1}] line was integrated after subtracting a smoothed H$\alpha$ profile, again assuming symmetry about 6563\,\AA~in the H$\alpha$ line.} \\
\hline
201   & 1.0             & 0.114  & 0.151  & 0.403                & 0.059    & 0.196    & 0.614 \\
207   & 1.0             & 0.140  & 0.147  & 0.369                & 0.059    & 0.222    & 0.621 \\
268   & 1.0             & 0.218  & 0.135  & 0.247                & 0.039    & 0.253    & 0.749 \\
\hline
\end{tabular}
Errors are difficult to estimate for these values, as line edges and continuum levels have been estimated
by eye.  However, care was taken to treat each line similarly. Measurement errors alone 
(determined through repeated measurements) are $\sim$\,5\%.
\end{minipage}
\end{table*}

There appears to be a weak blue continuum in the 600+\,day spectra of SN~2011dh.
\citet{Maund:2004fy}, in a very high S/N spectrum of SN~1993J taken $\sim$\,10\,yr after explosion,
were able to associate a blue continuum (and a detection of the Balmer absorption-line series) with a companion B supergiant,
thereby strongly supporting the binary nature of the SN and identifying the components
-- a K-giant progenitor and a B-giant companion.  In the spectrum of SN~2011dh above, however, we cannot attribute
the blue continuum to any stellar companion: fitting a Rayleigh-Jeans curve to the apparent continuum yields
best-fit temperatures much too hot for a stellar source.  The blue continuum in SN~2011dh is instead most likely a pseudocontinuum caused by
many blended lines.  In addition, our spectra are more noisy at the blue end, and the blue rise may be partially
caused by increased noise.  {\it HST} photometry taken near this time provides a slightly redder colour than synthetic
photometry from our spectrum: $F555W - F814W = 0.69 \pm 0.03$\,mag \citep[641\,days;][]{VanDyk:2013tp},
compared to $\sim$\,0.34\,mag from our spectrum (628$+$678\,days).
Thus, we tilt our spectrum to match the {\it HST} $F555W - F814W$ colour and re-examine the result
for evidence of a stellar companion. Our conclusion is essentially unchanged: even after tilting our spectrum, the
blue pseudocontinuum yields unreasonably hot best-fit blackbody temperatures.

Interestingly, there is a broad H$\alpha$ emission line in spectra of SN~2011dh through at least 334\,days,
similar to the emission seen in SN~1993J, \citep{Filippenko:1994is}, SN~2007Y \citep{Stritzinger:a},
and SN~2008ax \citep{Milisavljevic:2010cl} around the same time. 
There is also some indication of very broad H$\alpha$ in the spectra  of SN~2011dh at 600+\,days, though the S/N is low.
At late times the H$\alpha$ emission of SN~1993J was unambiguously identified with
interactions between the expanding SN shock wave and circumstellar material produced
by mass loss from the progenitor \citep[e.g.,][]{Patat:1995va,Houck:1996bs,Matheson:2000ff}. As we discuss
in \S\ref{sec:latespec}, SN~2011dh seems to present us with a more complex situation.

SN~2011dh, like SN~2001ig, displayed a relatively strong \Mg~line -- significantly
more prominent than \ion{Mg}{1}] in spectra of SN~2008ax \citep{Silverman:2009fo,Chornock:2011dj}.
This is especially apparent around day 334, where \ion{Mg}{1}] emission almost matches the emission in [\ion{Ca}{2}] and [\ion{O}{1}]. 
At very late times, in the 628$+$678\,day spectrum,
the \ion{Mg}{1}] is still quite apparent, though [\ion{O}{1}] and [\ion{Ca}{2}] have faded into the noise. Unfortunately, our 628 and 678\,day spectra
do not go much blueward of the \ion{Mg}{1}] emission peak; this, together with high noise levels at the blue end, prevents us from measuring the integrated flux reliably
at these times. The \ion{Na}{1}~D flux is also remarkably strong in the 600$+$ day spectra, as discussed below.
 
\subsection{The Spectrum of SN~2011dh at 600$+$ Days}
\label{sec:latespec}

\begin{figure*}
\centering
\includegraphics[width=\textwidth]{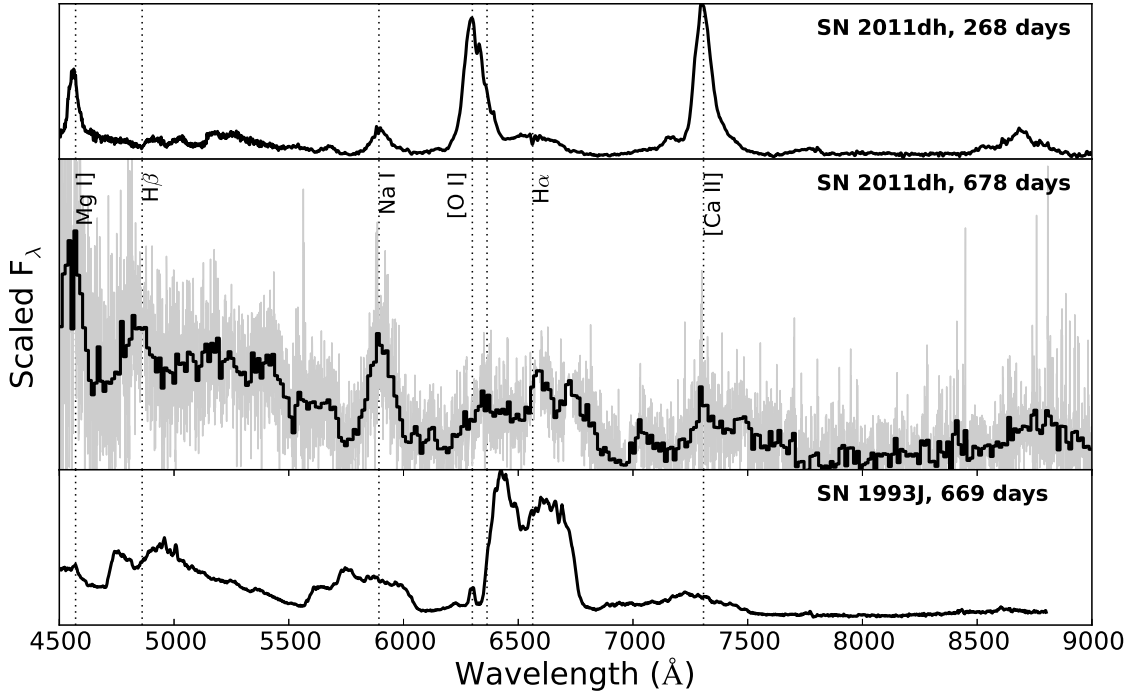}
\caption{SN~2011dh at 268 and 678 days, with SN~1993J at a similar epoch for comparison \citep[SN~1993J spectrum from][]{Matheson:2000ff}.
                The later spectrum of SN~2011dh has been rebinned to increase the S/N; the unbinned spectrum is shown in the background.
                In both SNe, the broad H$\alpha$ emission likely comes from interaction between an expanding shockwave and the circumstellar medium,
                but the \ion{Na}{1}~D line emission in SN~2011dh is most likely powered by $^{56}$Co decay while SN~1993J has a
                \ion{Na}{1}~D line powered by circumstellar interaction.}
\label{fig:latespec}
\end{figure*}

Recent photometry of the site of SN~2011dh taken by {\it HST} 
\citep[][]{VanDyk:2013tp}
and the Nordic Optical Telescope \citep[NOT;][]{Ergon:2013td}
provide late-time flux measurements of SN~2011dh.  Our latest two spectra, taken around the same time, confirm that
the optical flux measured by these groups was predominantly from the SN remnant and not, for example, from a binary companion to the
progenitor star.  678\,days after explosion, SN~2011dh continues to show clear \ion{Na}{1}~D emission with approximately the same
width as at earlier epochs (see Fig.~\ref{fig:latespec}). [\ion{Ca}{2}] emission is present but much reduced relative to \ion{Na}{1}~D, \ion{Mg}{1}] is 
still relatively strong but is buried in the noise at the blue end of the spectrum, and there is a very broad 
feature near $6500$ \AA~-- most likely a blend of broad H$\alpha$ and the [\ion{O}{1}] doublet. 

\citet{Ergon:2013td} report a slight fading of 2011dh between days 601 and 685: $0.009 \pm 0.0026$\,mag\,day$^{-1}$ in $V$.
The $V$-band decline rate between the last observation reported by \citet[$19.44 \pm 0.12$\,mag, 2012 Apr.~4, 310\,days]{Tsvetkov:2012ty}
and the first observation by \mbox{\citet[$22.56 \pm 0.10$\,mag, 2013 Jan.~20, 601\,days]{Ergon:2013td}}
is $\sim$\,0.011\,mag\,day$^{-1}$.  Both of these rates are notably less rapid than the $0.021$\,mag\,day$^{-1}$
decline rate measured from the 65--310\,day photometry by \citet[see \S\ref{sec:fluxcal}]{Tsvetkov:2012ty}; it seems that the flux decline rate is slowing down
at $t \gtrsim$\,300\,days.

\begin{figure}
\centering
\includegraphics[width=0.5\textwidth]{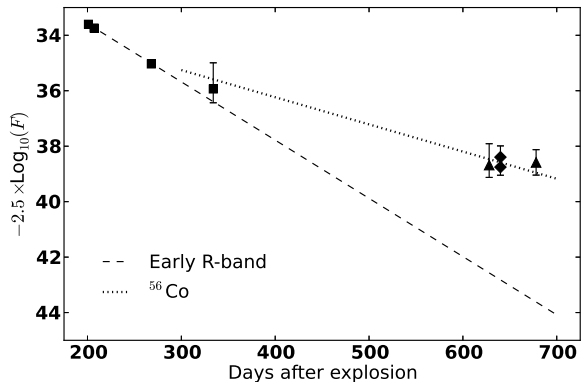}
\caption{Integrated \ion{Na}{1}~D flux in nebular spectra of SN~2011dh.  All values plotted here are presented in Table
               \ref{tab:nad}. The dashed line shows the early-time decline rate for the $R$ band (see \S\ref{sec:fluxcal})
               and the dotted line shows the decline rate of $^{56}$Co (0.0098\,mag\,day$^{-1}$).
               Boxes are used for spectra normalised to \citet{Tsvetkov:2012ty} photometry, triangles for spectra normalised to 
               \citet{Ergon:2013td} photometry, and diamonds for spectra normalised to \citet{VanDyk:2013tp} photometry.
               The positron-dominated epoch appears to begin somewhere between 300--350 days.
}
\label{fig:nad}
\end{figure}

\ion{Na}{1}~D provides the clearest feature in our 600$+$\,day spectra and is unambiguously associated with the SN (see Fig.~\ref{fig:latespec}).
We measured the integrated flux in the \ion{Na}{1}~D line for each of our spectra in the nebular phase; the results are shown in Table \ref{tab:nad} 
and Figure~\ref{fig:nad}.  Note that the absolute flux calibrations of long-slit spectra are often unreliable. As described in \S\ref{sec:fluxcal}, we address this by flux calibrating
our spectra to published photometry.  Quoted uncertainties include estimated error due to spectral noise, reported photometric errors, and
estimated measurement errors, all added in quadrature.  As Figure~\ref{fig:nad} illustrates, the \ion{Na}{1}~D line flux mirrors the trends in the photometry,
falling at the early $R$-band decline rate through $\sim$\,300\,days but then deviating significantly
around 300 or 350\,days and fading much more slowly through $\sim$\,600\,days.

This slowdown in the flux decay rate is possibly indicating a transition to $\gamma$-ray transparency in the ejecta of SN~2011dh.
As described in more detail by \citet{arnett:1996}, $^{56}$Co radioactivity ($^{56}$Co $\rightarrow$ $^{56}$Fe with a half-life of $\sim$\,77 days)
is the dominant source of energy for SNe at these epochs.  $^{56}$Co decay produces both $\gamma$-rays and high-energy
positrons.  The kinetic energy of the positrons is very likely to be deposited into the ejecta 
(and therefore contribute to the nebular line flux), while the fraction of $\gamma$-ray energy that gets deposited
depends upon the optical depth of the ejecta to $\gamma$-rays.  As the ejecta expand and the optical depth drops,
a larger fraction of the $\gamma$-rays escape, carrying their energy with them.

As the $\gamma$-ray energy deposition fraction drops, the SN fades away faster than the
$^{56}$Co rate (0.0098\,mag\,day$^{-1}$), until such time as the ejecta are transparent to $\gamma$-rays and approximately
all of them escape.  At that point, positron energy deposition
dominates the energy input of the ejecta and the bolometric flux decline rate is expected to follow the $^{56}$Co rate closely.
Broadband photometry should exhibit roughly the same behaviour. 
Assuming that the ejecta's abundance of neutral sodium is constant and that the exposure to heating does not
change significantly, so should the \ion{Na}{1}~D line flux.
The flux decline rates for both the late-time $V$-band photometry and the \ion{Na}{1}~D line flux in SN~2011dh
are consistent with a transition to a positron-powered ejecta sometime between 300 and 350\,days. 
Because positrons deposit their energy locally (near the decaying $^{56}$Co) but $\gamma$-rays deposit energy
throughout the ejecta, the transition to a positron-dominated energy input is likely to correspond to a change in the dominant
emission lines.  In SN~2011dh, we see that \ion{Na}{1}~D and \ion{Mg}{1}] emission stays strong in the positron-dominated
epoch while [\ion{O}{1}] and [\ion{Ca}{2}] fade away.  More detailed modeling is necessary to test this scenario, however.

\begin{table}
\begin{minipage}{.45\textwidth}
\caption{Photometric normalisations}
\label{tab:nad}
\begin{tabular}{ *5c }
\hline
Day\footnote{Days since explosion (2011 May 31.5).} & Passband & Mag $\pm 1\sigma$ &
\ion{Na}{1}~D flux\footnote{Units: $10^{-15}$\,erg\,s$^{-1}$\,cm$^{-2}$.} \\
\hline

201         & $R$     & $16.46\pm 0.06$ & $36.26\pm 2.10$ \\
207         & $R$     & $16.58\pm 0.07$ & $31.95\pm 2.24$ \\
268         & $R$     & $17.76\pm 0.10$ & $9.69\pm 1.14$ \\ 
334         & $R$     & $19.04\pm 0.14$ & $4.22\pm 2.45$ \\ 
628         & $V$     & $22.80\pm 0.26$ & $0.338\pm 0.171$ \\
628$+$678\footnote{To estimate the spectrum near the {\it HST} observations, we produce an average of
                              the 628 and 678\,day spectra.  We flux calibrate both spectra to $R = 0.0$ mag, coadd them,
                              and renormalise the result to the {\it HST} photometry.  This produces an equally weighted average
                              between the two spectra -- we assume the SN is changing slowly at this epoch and that
                              this averaged spectrum is a good measure of the relative flux near the average time (653\,days).}   &
                                     $F555W$ & $23.198\pm 0.019$ & $0.315\pm 0.098$ \\
628$+$678$^c$   & $F814W$ & $22.507\pm 0.022$ & $0.439\pm 0.137$ \\
678         & $V$     & $23.32\pm 0.38$ & $0.368\pm 0.243$ \\
\hline
\end{tabular}
Photometric normalisations applied to our nebular spectra and resultant absolute
flux in the \ion{Na}{1}~D line.  See \S\ref{sec:fluxcal} for a description of how we calculated 
the photometric estimates.  We measured the integrated line flux in the manner
described in \S\ref{sec:ana}, and the quoted errors include photometric normalisation errors,
spectral noise, and estimated measurement errors added in quadrature.
\end{minipage}
\end{table}

Note that the above discussion assumes the bolometric light curve is completely powered by $^{56}$Co.
Any additional energy input could be a confounding factor; most importantly, there may be a flux contribution from
shockwave interactions with circumstellar gas.
The progenitors of SNe~IIb are stars that have lost much of their hydrogen envelope, either through radiative winds
or through stripping by a binary companion.  If a significant amount of that material remains nearby in a cloud
of circumstellar matter, the expanding SN ejecta will impact it and form a shock boundary.
This shocked region produces high-energy photons which are then reprocessed down to optical wavelengths by material
in the outer shells of the ejecta, thereby producing broad emission lines and (possibly) a blue pseudocontinuum
\citep[for a more complete description of this process, see, e.g.,][]{Chevalier:1994ds,Fransson:1984a}. 
Late-time emission from circumstellar interaction (CSI) is common in SNe~IIn \citep[e.g.,][]{Fox:a}, which have lost a majority of their envelope
in the years prior to explosion, and it has been observed in other SNe~IIb \citep[e.g.,][]{Matheson:2000ff}. 

CSI could be augmenting the \ion{Na}{1}~D flux discussed above through either \ion{Na}{1}~D
or \ion{He}{1}\,$\lambda 5876$ emission.  As Figure~\ref{fig:latespec} shows, SN~1993J clearly displayed 
a shockwave-powered \ion{Na}{1}~D $+$ \ion{He}{1} blend with $\sim$\,1/3 the flux of the H$\alpha$ line \citep{Matheson:2000ff}. 
In the \citet{Chevalier:1994ds} model, approximately all of the shockwave-powered \ion{Na}{1}~D flux is emitted from a thin shell at the 
boundary between the unshocked circumstellar material and the shocked ejecta, while the H$\alpha$ flux comes from both the
thin shell and the shocked ejecta. This would imply a more boxy line profile for \ion{Na}{1}~D than H$\alpha$ (see below for further
discussion of the nebular H$\alpha$ line in SN~2011dh).
However, the late-time \ion{Na}{1}~D emission of SN~2011dh has a relatively narrow profile with no evidence of the box-like 
shape that would be expected if it were mostly CSI-powered, and the profile does not appear to change significantly between the
spectra taken at $<$\,1\,yr and those taken at $>$\,1\,yr.
It seems clear that, at 600$+$ days, the dominant source of \ion{Na}{1}~D flux remains radioactive decay and not CSI, though there may
well be some small amount of shockwave-powered \ion{He}{1} and \ion{Na}{1}~D emission buried in the noise.  If SN~2011dh's CSI-powered
\ion{Na}{1}~D $+$ \ion{He}{1} flux were a factor of 3 less than the H$\alpha$ flux (as was the case in SN~1993J), we would not expect to be able
to distinguish it from the noise in our spectra.

Note that the $V$-band data may, in addition, include flux contributions from a CSI-powered blue pseudocontinuum at these late times --
the 600$+$\,d spectra do indicate a faint continuum blueward of 6000\,\AA.  Because of the large time gap between our spectra at
334\,d and 628\,d, we do not know exactly when this blue pseudocontinuum emerged.  It does not seem to evolve
significantly between our spectra taken at 628 and 678\,days, and so we believe that the flux decline measured around this time by \citet{Ergon:2013td} 
($0.009 \pm 0.0026$\,mag\,day$^{-1}$) is dominated by the fading $^{56}$Co contribution and not by any evolving CSI-powered flux.
Note that $V$-band photometry likely also includes contributions from the \ion{Na}{1}~D, [\ion{O}{1}], [\ion{Fe}{2}], and H$\alpha$ lines, 
whether they are powered by radioactivity or CSI --- a rather complex puzzle to decode.
However, this complexity does not affect the measurement of individual line fluxes.  As described above, the \ion{Na}{1}~D line
indicates that the ejecta of SN~2011dh became fully $\gamma$-ray transparent (and therefore powered through positron energy deposition) 
between 300 and 350\,days after core collapse.

In contrast with the \ion{Na}{1}~D line, the nebular H$\alpha$ line appears to be fueled entirely through CSI.
Assuming the broad feature near 6563\,\AA~in
the 678\,day spectrum of SN~2011dh (see Fig.~\ref{fig:latespec}) is a broad and boxy H$\alpha$ feature,
it exhibits a full width at half-maximum intensity~(FWHM) of roughly 21,000--26,000\,\kms\ (there are large errors when
measuring the line width, as this spectrum has a low S/N and the line is weak).
Spectra taken the first month after core collapse show a blueshifted H$\alpha$ absorption component around 
15,400\,\kms\ to 12,500\,\kms\ \citep[velocities at 4 to 14\,days;][]{Marion:2013vr}.
These measurements mesh with the CSI scenario described above, wherein the unslowed outer ejecta impact the circumstellar material
and produce a shell of emitting gas that continues to move outward at its initial expansion velocity.

The scenario that SN~2011dh presents to us in its late-stage evolution is notably different than that of SN~1993J 
or SN~1987A.  As shown by \citet{Suntzeff:1992iy}, the peculiar Type II-P SN~1987A exhibited a continuously declining (yet nonzero) $\gamma$-ray opacity
until the slowly decaying isotope $^{57}$Co became the dominant source of energy around 800--900\,days after explosion
($^{57}$Co $\rightarrow$ $^{57}$Fe with a half-life of $\sim$\,272\,days).  This indicates that SN~1987A had a significantly higher $\gamma$-ray opacity than SN~2011dh.
In contrast, CSI became the dominant flux source in SN~1993J around 350\,days, when the spectrum became dominated
by broad and boxy emission lines \citep{Matheson:2000ff}, and it is impossible to tell when (or if) the radioactive energy deposition 
entered the $^{56}$Co positron-powered phase.  
Of course, several questions remain about SN~2011dh and its late-time evolution. Unless it rebrightens, however, the SN is 
too faint to hope for a significantly higher S/N spectrum than those presented here (our spectrum at 678\,days
represents an hour of integration time on a clear night with the 10\,m Keck II telescope).  Continued photometric
monitoring should provide more information as the SN evolves.

\subsection{The Oxygen Line Profile and the Ejecta Geometry}
\label{sec:geometry}

\begin{figure*}
\centering
\includegraphics[width=\textwidth]{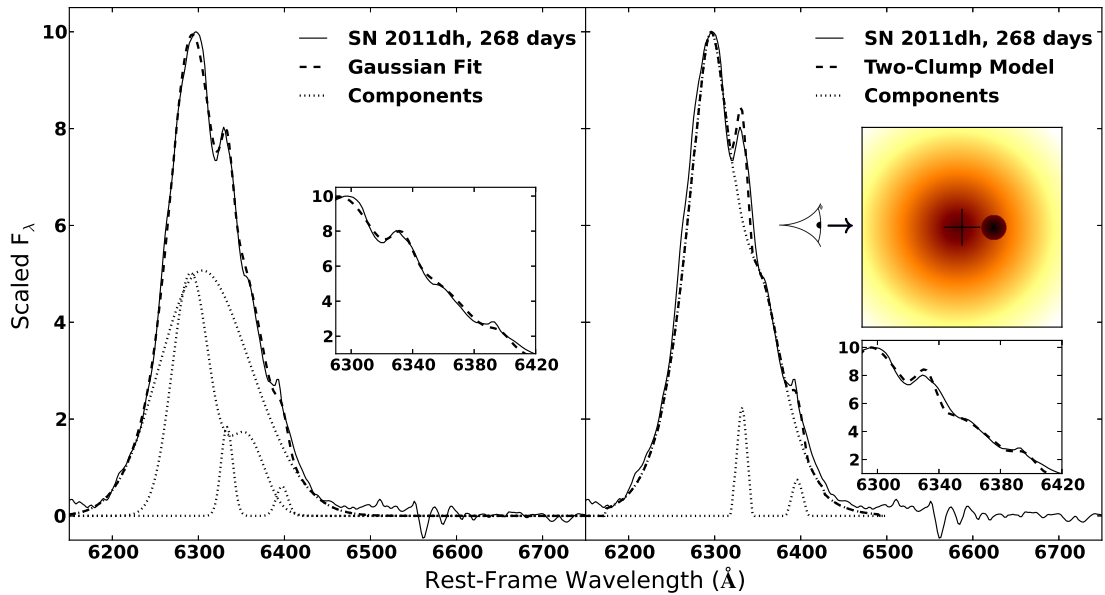}
\caption{The left panel shows a multi-Gaussian decomposition of SN~2011dh's \OI~doublet 268\,days after explosion. Each component is a doublet with a flux ratio
                of 3:1 and a wavelength separation of 64\,\AA.  Our best fit includes three such components: a broad central component, a large blueshifted
                clump, and a small redshifted clump.  The inset shows a magnified view of the profile's red side.
                The right panel shows the same observed line profile with our best-fit two-component model overlaid.
                The box on the upper right shows a crosscut through the emissivity profile in velocity space, 
                where the colour gradient represents log$_{10}$ of emissivity density and the cross
                marks the rest velocity of M51.  The box on the lower right is a magnified view of the right side of the
                line profile.  See \S\ref{sec:geometry} for a complete description. A colour version of this figure is available in the online journal.
}
\label{fig:geometry}
\end{figure*}

\begin{figure*}
\centering
\begin{subfigure}[b]{0.49\textwidth}
\centering
\includegraphics[width=\textwidth]{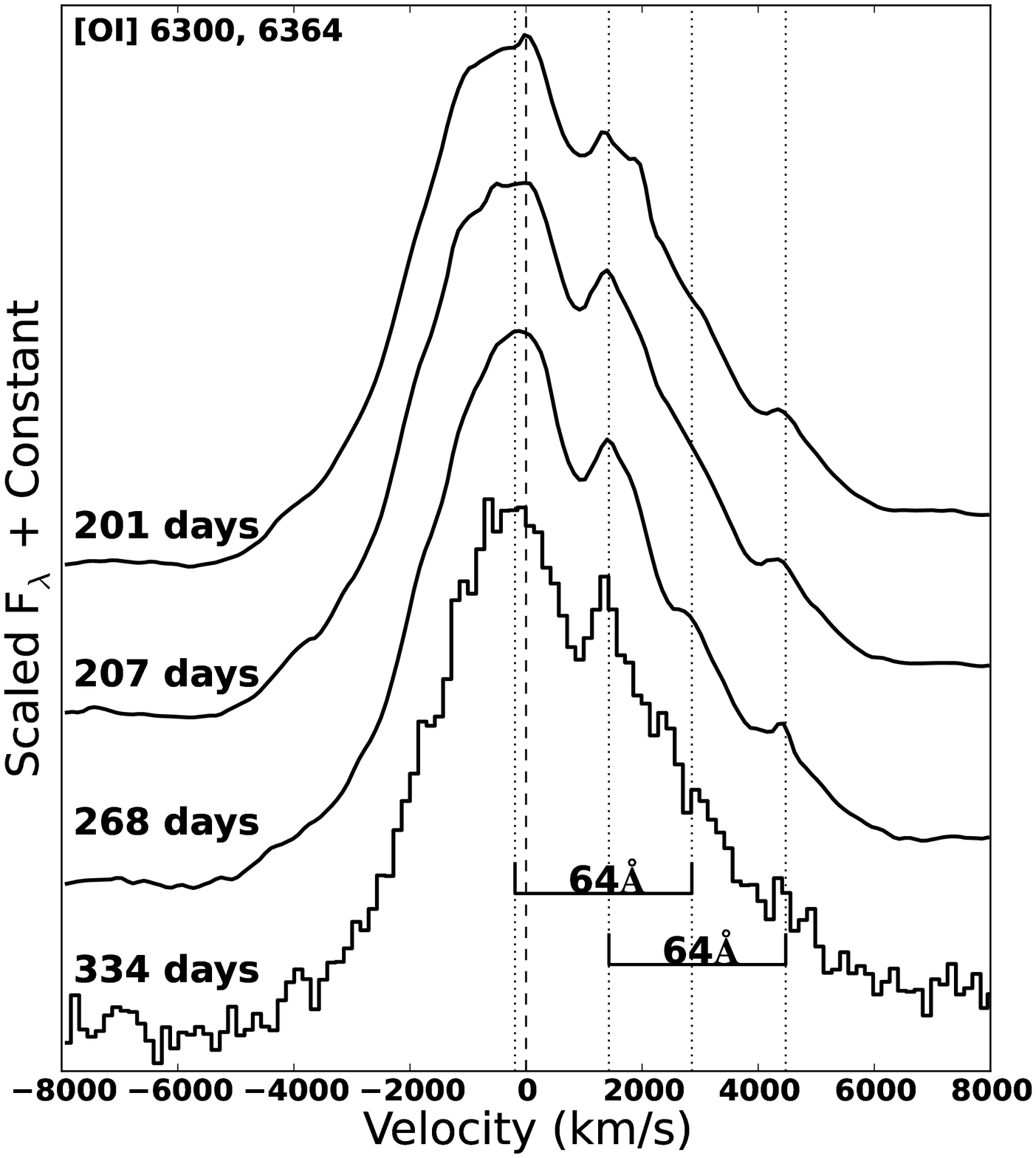}
\end{subfigure}
~
\begin{subfigure}[b]{0.49\textwidth}
\centering
\includegraphics[width=\textwidth]{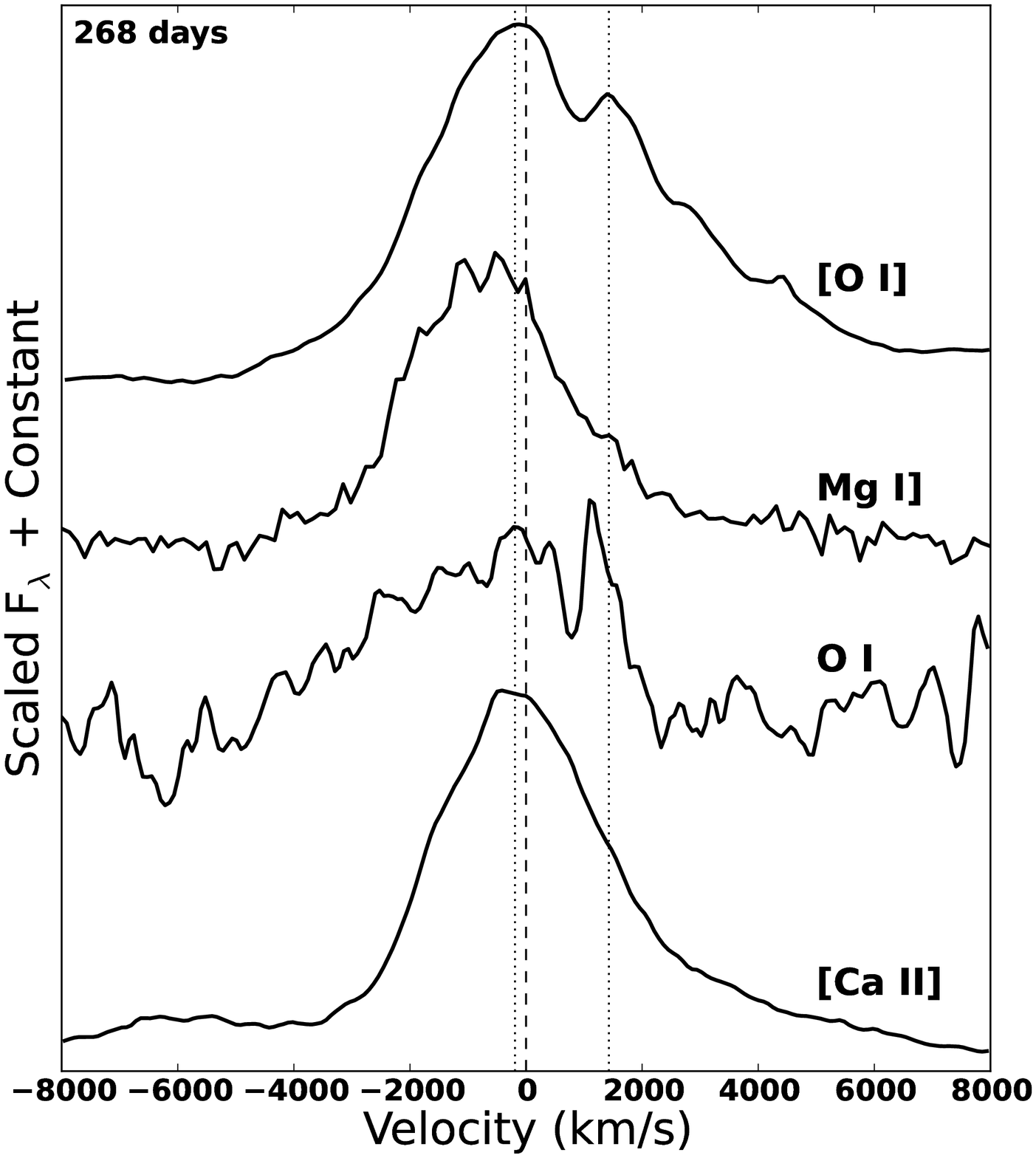}
\end{subfigure}
\caption{The \OI~doublet line profile during the early nebular phase of SN~2011dh (left) and the 
                \OI, \Mg, \ion{O}{1}\,$\lambda 7774$, and \Ca~profiles of SN~2011dh 268\,days after core collapse (right).
                The components described in \S\ref{sec:geometry}
                persist throughout the nebular phase with similar relative fluxes and wavelength offsets, and
                similar profiles are apparent in the [\ion{O}{1}], \ion{Mg}{1}], and \ion{O}{1} lines.
                The vertical dotted lines show the 
                best-fit velocities of the two components of the \OI~line as described in \S\ref{sec:geometry} and shown in Figure~\ref{fig:geometry}.
                The dashed line at 0\,\kms\ marks the rest frame of M51.
}
\label{fig:prof}
\end{figure*}

 \citet{Fransson:1989ie} showed that, given the reasonable assumptions of homologous expansion and 
optically thin emission, the profiles of forbidden lines in nebular SN spectra can be used as tracers of
the geometry and density profile of the emitting material.  The \OI~doublet, specifically, has been used as a
diagnostic of ejecta asphericity in Type Ibc/IIb SNe by many studies
\citep[e.g.,][]{Mazzali:c,Taubenberger:2009go, Milisavljevic:2010cl, Modjaz:2008uq, Maurer:2010hw, Maeda:2008dq}.
The [\ion{O}{1}] doublet is generally used because it is consistently one of the strongest lines in nebular SN spectra
and is largely isolated and unblended, and oxygen is one of the most abundant elements in stripped-envelope
core-collapse SNe.  The structure apparent in the [\ion{O}{1}] doublet has often been attributed to either 
a jet or torus geometry in the ejecta with the diversity of line profiles explained through viewing-angle
dependencies \citep[e.g.,][]{Mazzali:c,Maeda:2008dq, Modjaz:2008uq}, though other
explanations have been presented for some SNe \citep[e.g.,][]{Maurer:2010hw, Milisavljevic:2010cl}.

The [\ion{O}{1}] profile of SN~2011dh prominently displays multiple peaks and troughs.  We explore the geometrical implications
of this line profile in two ways.  Several studies have previously explored [\ion{O}{1}] line profiles by decomposing the profile
into a set of overlapping Gaussian curves, effectively assuming a multi-component Gaussian spatial distribution \citep[e.g.,][]{Taubenberger:2009go}.
We performed a similar fit for comparison, but
the spatial distribution of emissivity is not necessarily Gaussian; the choice to decompose the profile this way is mainly for convenience.
We also ran three-dimensional (3D) nebular radiative transfer models for a variety of geometries, attempting to fit the observed line profile
with a simple and physically plausible ejecta geometry.  
\OI~is a doublet with two peaks separated by 64\,\AA, with their relative intensities determined by the local density of neutral oxygen.
The intensity ratio reliably approaches 3:1 in nebular SN spectra \citep[e.g.,][]{Chugai:a,Li:a},
and we assume this holds true for SN~2011dh at these epochs.  Before analysing the [\ion{O}{1}] profile we remove the H$\alpha$ emission
by assuming it is symmetric about the rest wavelength (6563\,\AA) and subtracting a smoothed profile.
Note that this nebular spectral analysis is not a well-posed inverse problem;
many different geometries could produce the same spectral profile.

The results of our Gaussian decomposition of the line profile are shown in the left panel of Figure~\ref{fig:geometry}.  
For each component in our fit we specify the amplitude, position, and width of the 6300\,\AA~line; the properties of the 6364\,\AA~line are then forced.
In the spectrum at 268\,days past core collapse, our best fit to the [\ion{O}{1}] line requires three such components. 
There is a broad component blueshifted by $\sim$\,250\,\kms, a narrow component blueshifted by $\sim$\,400\,\kms, 
and a second narrow component redshifted by $\sim$\,1600\,\kms.  Note, however, that the broad component is only needed
because the overall line profile is distinctly non-Gaussian.  A more nuanced approach (below) provides a good fit to the profile with
only two components.

The results of our 3D modeling analysis are shown in the right panel of Figure~\ref{fig:geometry}.
In our models, we specify the emissivity of the [\ion{O}{1}] doublet in each spatial zone and integrate the transfer equation using the Sobolev approximation 
under the assumption that the ejecta are optically thin \citep[see, e.g.,][]{Jeffery:a}.  We decompose the 3D emission into multiple overlapping spherical
clumps, each with an exponentially-declining emissivity profile.
The primary peak is well fit by a sphere with an emissivity profile characterised by an exponential falloff with $e$-folding velocity $v_e = 950$\,\kms.  
To match the position of the peak, we need to offset the entire sphere from the origin toward the observer by $\sim$\,250\,\kms.  
The secondary peak is well fit by placing a second smaller spherical clump along the observer's line of sight but moving away at $\sim$\,1500\,\kms.
The emissivity profile of this second clump has an $e$-folding velocity of $v_e = 300$\,\kms\ which terminates at 600\,\kms. 
The integrated emission from the primary sphere is $\sim$\,24 times greater than the integrated emission of the smaller clump,
though the peak local emissivity of the smaller clump is a factor of $\sim$\,4 higher than that of the primary sphere.   

The right panel of Figure~\ref{fig:geometry} shows that this model does a decent job of fitting all features in the [\ion{O}{1}] profile. 
Though this simple model invokes only two components, the true ejecta geometry could in fact consist of 
multiple clumps of similar or smaller spatial dimensions.  This is because it is only the
larger inhomogeneities located along the line of sight that produce noticeable and well-separated features in the line profile.  

It is unclear whether the clump-like structures we infer from the [\ion{O}{1}] doublet correspond to inhomogeneities in the distribution of 
the oxygen itself or the $^{56}$Ni that excites it.  In 3D core collapse simulations, convective motions during neutrino heating act as 
seeds for Rayleigh-Taylor instabilities that develop when the shock passes through compositional interfaces \citep[e.g.,][]{Hammer:a}.
This results in fingers of heavier elements, such as $^{56}$Ni, punching out into the overlying layers of lighter elements.  Such a picture 
could explain the irregular line profiles seen in several core-collapse SNe at late times \citep[e.g.,][]{Filippenko:1985f,Matheson:2000ff}, and
has been explicitly considered previously for the asymmetry seen in SN~1987A. In particular, the substructure noted in the  H$\alpha$ line profile 
of SN~1987A \citep[the ``Bochum event'';][]{Hanuschik:a} has been interpreted as resulting from a relatively high velocity ($\sim$\,4700\,\kms)
``bullet" of $^{56}$Ni \citep{utrobin:a}.  A similar geometry could potentially 
be applicable to SN~2011dh, assuming a sizable (but slower) clump of $^{56}$Ni was mixed into the oxygen layer.  More sophisticated 3D 
nebular spectral modeling will be needed to constrain the geometry in more detail. For example, extending the secondary clump of our model
by adding more material to the extreme redshifted edge (making the clump aspherical) would fill in the discrepancies apparent
in the line profile near 6340\,\AA~and 6420\,\AA, and would also more closely resemble the extended structures apparent
in the models of \citet[][see their Fig.~2]{Hammer:a}.

Though the primary emitting sphere in our model is slightly offset from the origin, this
may be due to uncertainty in the true SN Doppler velocity rather than an actual asymmetry
of the ejecta.  Though M51 is almost face-on, \citet{Tully:a} show that the southeast quadrant of M51 (where SN~2011dh occurred) is
rotating toward us. The line-of-sight motion is significantly less than 250\,\kms\ but the true Doppler velocity of M51 may be lower
than the value we used to deredshift our spectra.  Adopting $z=0.00155 \pm 0.0002$ \citep{Falco:a} instead of the $z=0.002$ used in the rest
of this paper places the primary emitting component at a blueshift of $\sim$\,120\,\kms, within a factor of 2 of
the expected line-of-sight rotational velocity of M51 at the SN position. 
In addition, there are narrow H$\alpha$ lines from the host galaxy superimposed on our spectra (slight oversubtractions
are apparent in both the 268 and 334\,d spectra; see Figure~\ref{fig:compspec}).  Assuming the strongest of these lines in our 268\,d
spectrum was emitted in the rest frame of the SN, we measure a Doppler velocity of $550 \pm 160$\,\kms. Adopting this value places
the primary component at a blueshift of $\sim$\,200\,\kms\ relative to rest.
It appears that the primary emitting component of SN~2011dh is either symmetric with respect to the rest frame or nearly so.
Note, as well, that these relatively low Doppler velocities lie near the resolution limit of our spectra.  As described by \citet{Silverman:2012be},
our spectra have characteristic wavelength errors of 1 -- 2\,\AA~($\lesssim$90\,\kms).
The positions of most other nebular lines are consistent
with this scenario (with widths of several thousand \kms\ and irregular profiles, it is difficult to determine the centres of nebular
SN lines to high precision).  The \Mg~line is an exception, displaying a strong asymmetry and a blueshifted peak (see below).

As Figure~\ref{fig:prof} shows, the components described above persist from 201 to 334\,days,
and similar components at similar relative positions are apparent in the \ion{O}{1}\,$\lambda 7774$ profile
and the \Mg~profile, though the primary component appears to be more blueshifted in \ion{Mg}{1}].
The \Ca~profile, however, exhibits a simple and singly-peaked profile.
Other studies have shown that it is common for \ion{Mg}{1}] and [\ion{O}{1}] to display similarly asymmetric line profiles while [\ion{Ca}{2}] remains
relatively symmetric \citep[e.g.,][]{Modjaz:2008uq,Milisavljevic:2010cl}.

SN~2011dh's nebular [\ion{O}{1}] profile is not consistent with the often-proposed simple torus model of emitting material.
An emission trough due to an overall torus-like geometry of emitting material would fall at the rest wavelength
of the line (in the SN rest frame).  As Figure~\ref{fig:geometry} shows, SN~2011dh instead displays an emission peak at roughly the rest wavelength;
if the main trough in the line profile were associated with the centre of a torus, it would have to be offset from the rest frame of M51 by $\sim$\,1000\,\kms, 
an offset inconsistent with the centres of other nebular emission lines. To explore this further we ran several models similar to the two-clump model described 
above but with a toroidal component at various viewing angles, and we found no physically plausible toroidal geometries that matched the profile well.

\citet{Maurer:2010hw} showed that foreground H$\alpha$ absorption was a reasonable explanation for the double [\ion{O}{1}] peak
in SN~2008ax.  It is possible that foreground hydrogen absorption is also affecting the oxygen profile in SN~2011dh:
early-time spectra indicate a hydrogen expansion velocity of 12,500--15,400\,\kms\ \citep[velocities at 14 and 4\,days;][]{Marion:2013vr},
and the peak of [\ion{O}{1}] emission is $\sim$\,12,000\,\kms\ (265\,\AA) blueward of H$\alpha$.
However, explaining all three bumps in the profile through foreground absorption would require three well-placed hydrogen
overdensities, at $\sim$\,8100, 9600, and 11,100\,\kms, and would not account for the line profiles of \Mg~and \ion{O}{1}\,$\lambda 7774$.
It seems apparent that the [\ion{O}{1}] line-profile asymmetries in SN~2011dh come from distinct emitting components moving relative to each other,
each displaying the doublet nature of the line.  Additionally, the lack of obvious H$\alpha$ absorption features may indicate a very low hydrogen shell mass.

\section{Nebular Models}
\label{sec:models}

\begin{figure*}
\centering
\includegraphics[width=\textwidth]{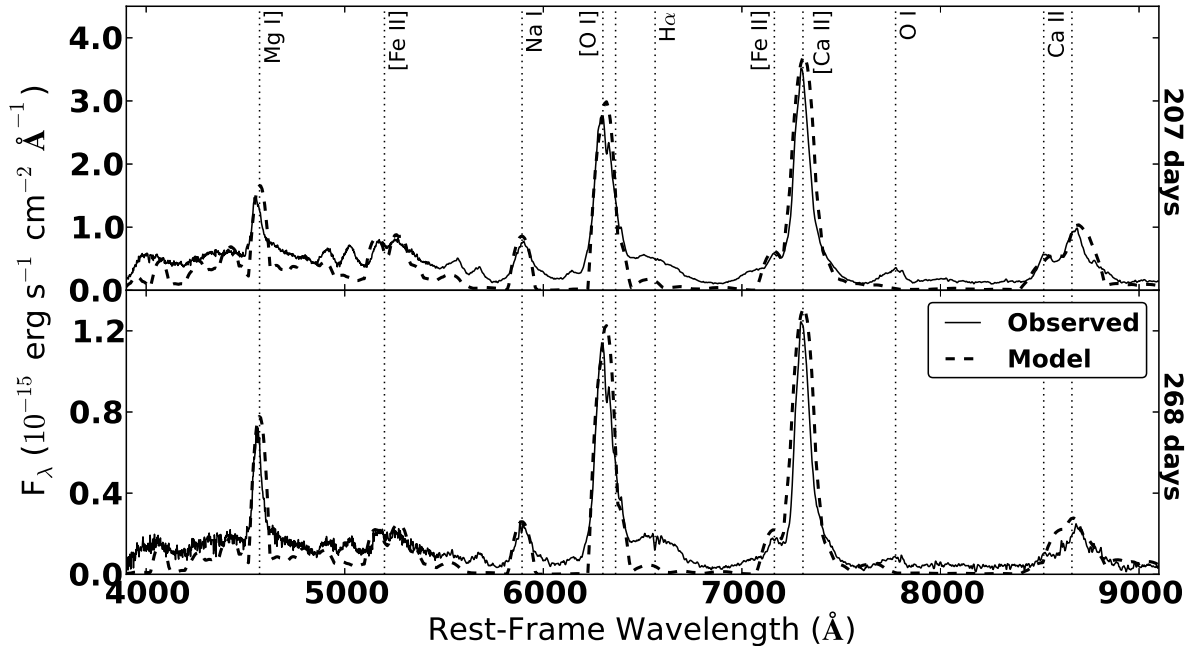}
\caption{Comparison between our best-fit nebular models and the observed spectra of SN~2011dh at 207 and 268 days 
                after core collapse.  See \S\ref{sec:models} for details.}
\label{fig:models}
\end{figure*}

We use a spherically symmetric single-zone non-LTE (local thermodynamic equilibrium) nebular modeling code to further explore
SN~2011dh.  The code tracks the heating of the nebular ejecta through
deposition of $\gamma$-rays and positrons produced by radioactive decay.  This heating is balanced
by line emission to determine both the temperature and ionization state of the nebula.  Following
methods and ideas first outlined by \citet{Axelrod:a} and \citet{Ruiz-Lapuente:a}, the code was developed by \citet{Mazzali:a},
and has been described in greater detail by, for example, \citet{Mazzali:2010ha} and \citet{Mazzali:b}.

The code is available in a one-zone version, a stratified version, and
a three-dimensional version. The one-zone model provides a rough estimate of
the properties of a SN nebular spectrum (e.g., mass and elemental abundances). The stratified model
is preferred when comparing a detailed model of the explosion with the data
\citep{Mazzali:2007dt}, and the three-dimensional model is useful for strongly asymmetric events
\citep{Mazzali:c}. Here, since the profiles of the emission lines do not
deviate significantly from the theoretically expected parabolic profiles and
developing a complete explosion model is beyond the 
scope of this paper, we restrict ourselves to the
one-zone approach. Mass-estimate differences between the one-zone
and the stratified model are relatively small in cases where the ejecta do not display strong asphericity
\citep[$\sim 20$\%;][]{Mazzali:a}. The code does not
include recombination emission, and therefore neither H$\alpha$ nor the \ion{O}{1}\,$\lambda 7774$
recombination line are reproduced. While hydrogen is located outside the carbon-oxygen core
and ignoring H$\alpha$ does not affect our result, ignoring the oxygen
recombination line can introduce an error, though it should be small \citep{Maurer:2010hw}. 
Another element of uncertainty is introduced by the fact that silicon does not
have strong lines in the optical range. The strongest line, [\ion{Si}{1}]\,$\lambda 6527$,
is about one third as strong as \ion{Na}{1}~D and is swamped by the H$\alpha$ emission.
All these effects could lead to an overestimate of the mass.
Finally, a major source of uncertainty is the subtraction of the background
pseudocontinuum.

Our best-fit models to the day 207 and 268 spectra are shown in Figure~\ref{fig:models}.
The spectrum of SN~2011dh changes only slightly between these two epochs, and the
two (independent) models are very similar.
These models are powered by $\sim$\,0.07\,M$_{\sun}$ of $^{56}$Ni and exhibit an outer envelope
velocity of $\sim$\,3500\,\kms\ with a total enclosed mass of $\sim$\,0.75\,M$_{\sun}$.
See Table \ref{tab:models} for a detailed listing of the mass composition of the models.
Note that these values are sensitive to errors in the determination of the distance
to SN~2011dh and errors in the absolute flux calibration of our spectra.
Most of the major features of these spectra are matched by the models, including
the prominent \Ca, \OI, \ion{Na}{1}~D, and \Mg~lines.

\citet{Bersten:2012ic} derived a similar but slightly lower $^{56}$Ni mass of $\sim$\,0.065\,M$_{\sun}$
from bolometric light curve modeling through the first 80 days, while \citet{Sahu:2013vd} derived
a slightly higher mass of $\sim$\,0.09\,M$_{\sun}$ through an analytic treatment of the bolometric light curve peak.
Several other SNe~IIb have been modeled in their nebular phase with similar codes, providing a 
useful set for comparison.  Our models of SN~2011dh include $\sim$\,0.26\, M$_{\sun}$ of oxygen, much less than was needed 
for SNe~2008ax, 2001ig, and 2003bg 
\citep[$\sim$\,0.51, 0.81, and 1.3\,M$_{\sun}$, respectively;][]{Maurer:2010hw,Silverman:2009fo,Mazzali:2009dd}.
The $^{56}$Ni mass required is also relatively low.  SN~2011dh had $\sim$\,0.067\,M$_{\sun}$ of nickel, but as the above authors
have shown, SNe~2008ax, 2001ig, and 2003bg required $\sim$\,0.10, 0.13, and 0.17\,M$_{\sun}$, respectively.
This indicates a relatively low-mass progenitor for SN~2011dh.

On the other hand, it is unclear whether the nebular spectra capture all of
the carbon-oxygen core. The lowest velocity of the He lines is $\sim$5000\,\kms, while the
width of the emission lines is only $\sim$3500\,\kms. Material between these
two velocities may not be captured by the nebular modelling. Although this would
go in the direction of compensating for the overestimate described above, we
conservatively assign an error of $\sim \pm 50$\% in our mass estimates.   It
would be interesting to test the data against realistic explosion models in
order to narrow down this uncertainty. This will be the subject of future work.

Multiple groups have modeled the nucleosynthetic yields of core-collapse SNe of various zero-age main sequence masses
 \citep[e.g.,][]{Woosley:a,Thielemann:a,Nomoto:2006di}.  Though there are some discrepancies between our nebular model 
and the nucleosynthetic models (and some disagreements between different nucleosynthetic modeling efforts),
our models are most consistent with a progenitor mass of 13--15\,M$_{\sun}$.   For example, \citet{Thielemann:a}
predict carbon yields of 0.06, 0.08, and 0.115\,M$_{\sun}$ and oxygen yields of 0.218, 0.433, and 1.48\,M$_{\sun}$ for 13, 15, and 20\,M$_{\sun}$ progenitors, respectively.
The values required by our best-fit model, 0.07\,M$_{\sun}$ of carbon and 0.26\,M$_{\sun}$ of oxygen, indicate a 13--15\,M$_{\sun}$ progenitor.
Note that not all elements are in such good agreement.

This result corroborates the findings of several other groups: the progenitor of SN~2011dh was a relatively low-mass ($\sim$\,13--17\,M$_{\sun}$) yellow supergiant
that likely had its outer envelope stripped away by a binary companion \citep[e.g.,][]{Bersten:2012ic,Benvenuto:2012gj,Maund:2011ez,VanDyk:2013tp,Murphy:2011em}.
SN~2011dh has provided a powerful test of the accuracy of SN progenitor studies through nebular spectra.
The clear agreement between the nebular modeling and the results of such varied
studies indicates that models of nebular SN spectra provide real and powerful constraints of the progenitor's properties. 
There is, however, work to be done to understand the discrepancies between modeled nucleosynthetic yields and nebular-spectra models.

\begin{table}
\caption{Nebular model mass composition}
\label{tab:models}
\begin{tabular}{ *3c }
\hline
Element & Mass (M$_{\sun}$) Day 207 & Mass (M$_{\sun}$) Day 268 \\
\hline
C    & $7.0 \times 10^{-2}$ & $6.0 \times 10^{-2}$ \\
O    & $2.6 \times 10^{-1}$ & $2.8 \times 10^{-1}$ \\
Na & $1.3 \times 10^{-4}$ & $1.7 \times 10^{-4}$ \\
Mg & $1.6 \times 10^{-3}$ & $4.2 \times 10^{-3}$ \\
Si  &  $3.0 \times 10^{-1}$ & $3.0 \times 10^{-1}$ \\
S   &  $2.3 \times 10^{-2}$ & $2.5 \times 10^{-2}$ \\
Ca & $9.1 \times 10^{-3}$ & $1.0 \times 10^{-2}$ \\
Fe  & $1.0 \times 10^{-2}$ & $1.0 \times 10^{-2}$ \\
$^{56}$Ni & $6.7 \times 10^{-2}$ & $7.0 \times 10^{-2}$ \\
Total          & $7.4 \times 10^{-1}$ & $7.6 \times 10^{-1}$ \\
\hline
\end{tabular}
Mass composition of non-LTE nebular models fit to spectra of SN~2011dh at 207 and 268\,days after core collapse.
\end{table}

\section{Conclusions}
\label{sec:conc}

SN~2011dh was a very nearby SN~IIb discovered in M51 in early June 2011, providing observers
with a valuable opportunity to track the evolution of one of these relatively rare SNe in detail.  The nature of SN~2011dh's progenitor star
has been much debated.  In this paper, we present nebular spectra
from 201 to 678\,days after explosion as well as new modeling results. 
We confirm that the progenitor of SN~2011dh was a star with a zero-age main sequence mass of 13--15\,M$_{\sun}$,
in agreement with the photometric identification of a candidate YSG progenitor.

In addition, our spectra at $\sim$\,2\,yr show that photometric observations taken near that
time are dominated by the fading SN and not, for example, by a background source or a binary companion.
We present evidence pointing toward interaction between the expanding SN blastwave 
and a circumstellar medium, and
show that the SN enters the positron-dominated phase by $\sim$\,1\,yr after explosion.
Finally, we explore the geometry of the ejecta through the nebular line profiles at day 268,
concluding that the ejecta are well fit by a globally spherical model with dense aspherical components or clumps.
In addition to the data presented here we have obtained several epochs
of spectropolarimetry of SN~2011dh as it evolved. The analysis of those data is beyond the scope of this paper,
but they will provide additional constraints on any asymmetry in the explosion of SN~2011dh.

\section*{Acknowledgements}
Sincere thanks to all of the supernova experts who contributed through discussions, including (but not limited to)
Brad Tucker, WeiKang Zheng, Ori Fox, Patrick Kelly, and J.~Craig Wheeler (whose keen eye identified a significant 
typo in the manuscript).  We thank the referee for suggestions that helped to 
improve this paper.
Some of the data presented herein were obtained at the W.~M.~Keck Observatory, which is operated as a 
scientific partnership among the California Institute of Technology, the University of California, and the 
National Aeronautics and Space Administration (NASA); the observatory was made possible by the generous 
financial support of the W.~M.~Keck Foundation. The authors wish to recognise and acknowledge the 
very significant cultural role and reverence that the summit of Mauna Kea has always had within the 
indigenous Hawaiian community; we are most fortunate to have the opportunity to conduct observations from this mountain.

This material is partially based upon work supported by a National Science Foundation (NSF) Graduate Research Fellowship
to J.B.~under Grant No.~DGE 1106400.  
J.M.S.~is supported by an NSF Astronomy and Astrophysics Postdoctoral Fellowship under award AST-1302771.
A.V.F.~and his SN group at UC Berkeley acknowledge generous support from
Gary and Cynthia Bengier, the Richard and Rhoda Goldman Fund, the
Christopher R.~Redlich Fund, the TABASGO Foundation, and NSF
grant AST-1211916.
This research has made use of NASA's Astrophysics Data System Bibliographic Services, as well as 
the NASA/IPAC Extragalactic Database (NED) which is operated by the
Jet Propulsion Laboratory, California Institute of Technology, under contract with NASA.

\bibliographystyle{mn2e}
\bibliography{xtra_bib,snae_bib}
\end{document}